# NMR and μ⁺SR detection of unconventional spin dynamics in Er(trensal) and Dy(trensal) molecular magnets

E. Lucaccini[1], L. Sorace[1,†], F. Adelnia[2,*], S. Sanna[3], P. Arosio[4], M. Mariani[2,‡], S. Carretta[5], Z. Salman[6], F. Borsa[2,7], A. Lascialfari[4,2]

[1] Dipartimento di Chimica "U. Schiff" and INSTM RU, Università degli studi di Firenze, Via della Lastruccia 3, 50019 Sesto F.no (FI), Italy

[2] Dipartimento di Fisica and INSTM, Università degli studi di Pavia, Pavia, Italy

[3] Dipartimento di Fisica e Astronomia, Università degli studi di Bologna, Bologna, Italy

[4] Dipartimento di Fisica "A. Pontremoli" and INSTM, Università degli Studi di Milano, Milano, Italy

[5] Dipartimento di Scienze Matematiche, Fisiche e Informatiche and INSTM, Università degli studi di Parma, Parma, Italy

[6] Laboratory for Muon Spin Spectroscopy, Paul Scherrer Institute, CH-5232 Villigen PSI, Switzerland

[7] Department of Physics and Ames Laboratory, Iowa State University, Ames, IA, USA

† Corresponding author: lorenzo.sorace@unifi.it

‡ Corresponding author: manuel.mariani@unipv.it

* Currently at Vanderbilt University, Institute of Imaging Science, Nashville , USA

**Abstract**

Measurements of proton Nuclear Magnetic Resonance ($^1$H NMR) spectra and relaxation and of Muon Spin Relaxation ($\mu^+$SR) have been performed as a function of temperature and external magnetic field on two isostructural lanthanide complexes, Er(trensal) and Dy(trensal) (where H$_3$trensal=2,2',2''-tris-(salicylideneimino)triethylamine) featuring crystallographically imposed trigonal symmetry. Both the nuclear $1/T_1$ and muon $\lambda$ longitudinal relaxation rates, LRR, exhibit a peak for temperatures T<30K, associated to the slowing down of the spin dynamics, and the width of the NMR absorption spectra starts to increase significantly at T~50K, a temperature sizably higher than the one of the LRR peaks. The LRR peaks have a field and temperature dependence different from those previously reported for all Molecular Nanomagnets.. They do not follow the Bloembergen-Purcell-Pound scaling of the amplitude and position in temperature and field and thus cannot be explained in terms of a single dominating correlation time $\tau_c$ determined by the spin slowing down at low temperature. Further, for T<50K the spectral width does not follow the temperature behavior of the magnetic susceptibility $\chi$. We suggest, using simple qualitative considerations, that the observed behavior is due to a combination of two different relaxation processes characterized by the correlation times $\tau_{LT}$ and $\tau_{HT}$, dominating for T<30K and T>50K, respectively. Finally, the observed flattening of LRR for T<5K is suggested to have a quantum origin.

# I. Introduction

Molecular Nanomagnets are characterized by regular crystalline structures in which the cores of adjacent molecules, containing a few exchange-coupled transition metal ions, are well separated by shells of organic ligands [1]. Hence the crystal behaves as an ensemble of identical and almost non-interacting zero-dimensional magnetic units, whose quantum behavior can be evidenced by macroscopic bulk measurements. Such molecules are of great interest for fundamental physics as model systems for the study of a variety of quantum phenomena, such as quantum-tunneling of the magnetization [2, 3, 4], Néel-vector tunneling [5], quantum entanglement between distinct, spatially separated cores [6-11], and decoherence [1, 12, 13].

Among these systems, a specific class of complexes is that of Single Molecule Magnets (SMMs), which feature a slow magnetic dynamics and magnetic hysteresis of purely molecular origin on long timescales [14]. This behavior is due to the presence of a magnetization reversal barrier arising as a consequence of a large spin ground state and an easy-axis type magnetic anisotropy, resulting in an Arrhenius - type dependence of the magnetization relaxation rate with temperature. The discovery of this behavior has opened new and interesting perspectives also for the potential technological applications of these molecules [15]. Indeed, it paves the way to build high-density magnetic memories by encoding a bit of information in each molecule: in this perspective, large efforts have been devoted to increase the size of the magnetic anisotropy barrier and thus the temperature at which magnetic bistability is observed on reasonable timescales [16, 17].

A seminal report by Ishikawa [18] showed that also molecules containing a single lanthanide ion can display slow relaxation of the magnetization at low temperature. These mononuclear lanthanide complexes, usually identified as single-ion magnets (SIMs), are particularly appealing for the possible realization of single-spin based storage devices, even at the atomic level [19]. Furthermore, the large magnetic moment and crystal field anisotropy of many Ln(III) ions results in magnetization reversal barriers much higher than in polynuclear clusters based on *3d* metal ions, opening up the possibility of magnetic data storage in single molecules at temperatures above liquid nitrogen [20-21]. It is, however, now well established that the blocking temperature (conventionally defined as the temperature at which magnetization relaxation time equals 100 s) [22] does not necessarily increase by increasing the barrier. This is essentially due to the presence of additional magnetization relaxation pathways, each of which shows a specific field and temperature dependence and has to be controlled if complexes with improved performance are sought for. Among these additional pathways, Quantum Tunneling of Magnetization is of paramount importance, since it hampers bistability in zero field and thus potential applications. In Kramers' ions lanthanide-based complexes this is usually attributed to hyperfine coupling to magnetic nuclei and dipolar fields from neighbouring molecules and it is of particular relevance for systems with low axiality of the magnetic anisotropy tensor. In addition, the large-energy Orbach steps are assisted by molecule-specific optical phonons rather than by simple Debye acoustic ones. Finally, Raman type relaxation mechanisms also appears to be much more important than in polynuclear *3d* molecules.

It is then clear that to unravel and pinpoint the nature and the role of the various mechanisms driving spin dynamics in these systems, a multi-technique approach on different timescales is necessary. In this respect, local spectroscopic techniques like Nuclear Magnetic Resonance (NMR) and Muon Spin Relaxation ($\mu^+$SR), which have been proved [23] to be useful and powerful probes of the spin dynamics in *3d* polynuclear complexes, appear much underused in this field [24, 25].
In all the SMMs investigated to date, the NMR and $\mu^+$SR longitudinal relaxation rate (LRR) have shown a maximum which occurs at a temperature where the frequency of the magnetic fluctuations slows down to a value close to the Larmor frequency of the nucleus or muon. All models successfully employed to analyze the data rely essentially on the assumption of a Lorentzian shape of the spectral density of the electronic spin fluctuations, and predict a universal scaling in amplitude and position of the peak vs. temperature and external magnetic field (i.e. Larmor frequency) [23, 26-29]. More precisely, this peak scales to lower values and displaces toward higher temperatures when the field is increased. In particular, the Bloembergen, Purcell and Pound (BPP) model [30,31] is based on the reasonable assumptions that the LRR is proportional to the effective magnetic moment $\chi T$, and that the geometric part of the hyperfine interaction between the nucleus/muon (for NMR/$\mu^+$SR respectively) and the magnetic ion, as well as the spin dynamical parameters, are independent of the applied magnetic field. The same scenario was recently found to apply also to Tb(III), Dy(III) [24] and Er(III) [32-34] based SIMs.

In this paper we present a combined $\mu^+$SR and NMR investigation performed on two isostructural lanthanide complexes, Er(trensal) and Dy(trensal) (where H$_3$trensal=2,2',2''-tris-(salicylideneimino)triethylamine) featuring crystallographically imposed trigonal symmetry [35] (see Fig. 1). The Ln(trensal) family has been widely studied, starting from luminescence experiments [33a, 34] until the more recent magnetic investigations [37, 38, 39]. These investigations showed that the trigonal crystal field introduces for both Er(trensal) and Dy(trensal) a large splitting of the 8 Kramers doublets of the J = 15/2 ground states, with eigenfunctions which are linear combinations of different M$_J$ values. The gap between the ground and first excited state in the two systems is about 62 K for Dy and 77 K for Er derivative [39]. Furthermore, the two complexes show different type of magnetic anisotropy in their ground state, namely easy axis for Er(trensal) and easy plane for Dy(trensal) [37, 40]. Despite the different anisotropy, slow relaxation of the magnetization was observed in applied field for both systems, demonstrating that the relaxation of the magnetization in the conditions used for alternated current (ac) susceptometry is not proceeding simply by thermally activated spin reversal over the anisotropy barrier. Rather, this was rationalized using a combination of direct, Raman and Quantum Tunneling processes. In this respect, the extremely detailed picture of the energy levels' structure of these complexes and the peculiar dynamics observed by ac susceptometry, make these systems ideal testing grounds for the application of NMR and $\mu^+$SR spectroscopy to the investigation of spin dynamics of lanthanide-based complexes.

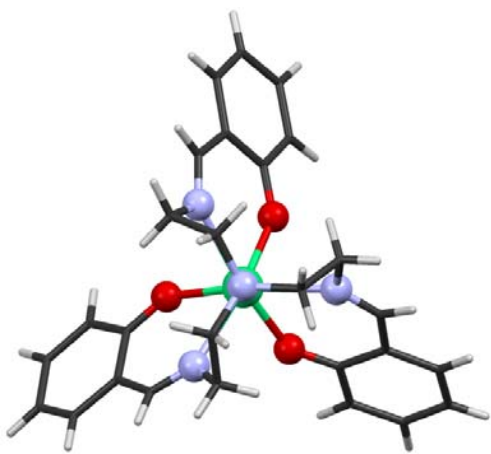

Fig. 1 (color online) Molecular structure, viewed along the trigonal axis, of Ln(trensal) [Ln=Dy, Er; H$_3$trensal=2,2',2''-tris-(salicylideneimino)triethylamine]. Color code: Lanthanide (green ball), oxygen (red ball), nitrogen (violet ball), carbon (black stick), hydrogen atoms (white stick).

The experimental results we report in the following cannot be justified within a BPP framework and, thus, have been tentatively explained by means of a qualitative phenomenological model, leaving to future investigations the development of a theoretical framework. In this model we assumed two different temperature ranges of spin dynamics, one at high temperature T > 50K, characterized by $\tau_{HT}$ and corresponding to a slow spin relaxation of magnetic excited states, and one at lower temperature (T < 50 K) characterized by $\tau_{LT}$. Furthermore, at even lower temperature (T< 5 K) the relaxation of the magnetization of the two complexes becomes temperature independent, thus suggesting the presence of a dominating quantum dynamical process.

## II. Experimental details

Dy(trensal) and Er(trensal) were prepared as reported elsewhere [35]. The crystallographic phase and purity of the sample has been checked by Powder X-ray diffractometry. Measurements were perfomed with a Bruker D8 Advance powder diffractometer equipped with a Cu source (K$\alpha$, $\lambda$ = 1.54 Å).

Magnetic DC susceptibility was measured on the two complexes in the form of powders on a MPMS-XL7 Quantum Design superconducting quantum interference device (SQUID) magnetometer in the temperature range 2−300 K at several applied magnetic fields, varying from 0.005 to 1.5 T.

NMR measurements were performed, by means of FT - pulse spectrometers, in the temperature range 1.5 < T < 300 K, at three different static applied magnetic fields ($\mu_0 H$ = 0.5, 1.5, 6.18 T). In particular the proton NMR spectra were obtained in two different ways: (i) for narrow lines the intensity of the radio-frequency pulse was sufficiently strong to irradiate the entire NMR spectrum, and thus the spectra were obtained from the Fourier transform (FT) of the half echo signal collected by applying the standard Hahn spin-echo pulse sequence; (ii) for broad lines, the line shape was obtained by plotting the envelope of the FTs of the echo signal by sweeping the frequency and keeping constant the applied magnetic field. We measured the $^1$H spin – lattice relaxation time $T_1$ through a spin echo saturation recovery sequence with a comb of 10 saturation pulses, preceding the spin-echo sequence for the signal detection; the recovery curves were obtained from the integration, through a homemade software, of the area under the echo signal as a function of saturation times (delay times) between the end of the comb pulses and the reading sequence. The $\pi/2$ pulse used was in the range 1.5 μs < $\pi/2$ < 4 μs (depending on the applied magnetic field).

$\mu^+$SR data were collected at the Paul Scherrer Institut (PSI, Villigen, Switzerland) large scale facility on GPS (for Dy(trensal)) and Dolly (for Er(trensal)) spectrometers. In both cases, three different longitudinal magnetic fields were applied ($\mu_0 H$ = 0.03, 0.1, 0.25 T) in the temperature range 2 - 200K.

## III. Magnetic susceptibility results

The results of magnetic susceptibility measurements at different applied magnetic fields are reported as $\chi_{mol}T$ (actually $M_{mol}\cdot T/\mu_0 H$) as a function of temperature in Fig. 2 (data at different fields were previously reported in ref. [37]). At room temperature the experimental values approach the limiting value corresponding to the free ion value for the $^4I_{15/2}$ and $^6H_{15/2}$ multiplets of $Er^{III}$ and $Dy^{III}$ (11.48 and 14.17 emu K mol$^{-1}$ respectively [41]); the decrease of the effective magnetic moment observed on lowering temperature is due to the progressive depopulation of the excited sublevels of the J=15/2 multiplets.

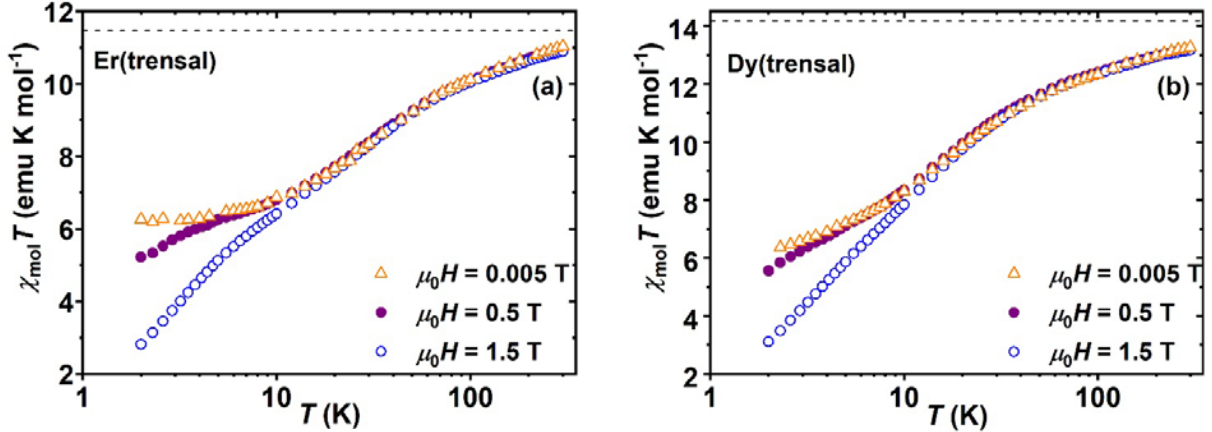

**Fig. 2** (color online) Temperature dependence of the product of the molar magnetic susceptibility with temperature, obtained as $\chi_{mol}T = (M_{mol} \cdot T)/(\mu_0 H)$, at different applied fields for Er(trensal) (a) and Dy(trensal) (b) samples. The dashed lines represent the free ion limit values expected for the two ions.

In particular, on the basis of the electronic structure of these systems, [36,37] we can attribute the relevant decrease of $\chi_{mol}T$ below 50-60 K to the beginning of the exclusive population of the doublet ground states of the two molecules. The onset of the field dependence of the $\chi_{mol}T$ values below 10 K is due to saturation effects, implying that $k_B T$ is not much larger than $g\beta\mu_0 H$ (here $\beta$ is the Bohr Magneton).

### IV. Proton NMR spectra

The proton NMR absorption spectra of both samples show a narrow central signal coming from protons far away from the magnetic RE ion and a broader base due to the distribution of local magnetic fields generated at the closest proton sites by the nuclear–electron (hyperfine) dipolar and contact interactions (Fig. 3). The width of the spectrum is then due to the large number of inequivalent protons, that sense slightly different magnetic fields. The width of the broad base spectrum increases clearly by lowering the temperature below T~90 K, and the whole shape of the spectrum appears determined by the onset of static local fields, suggesting at least a partial freezing of the RE magnetic moment on the NMR energy absorption timescale (some hundreds of kHz/few MHz).

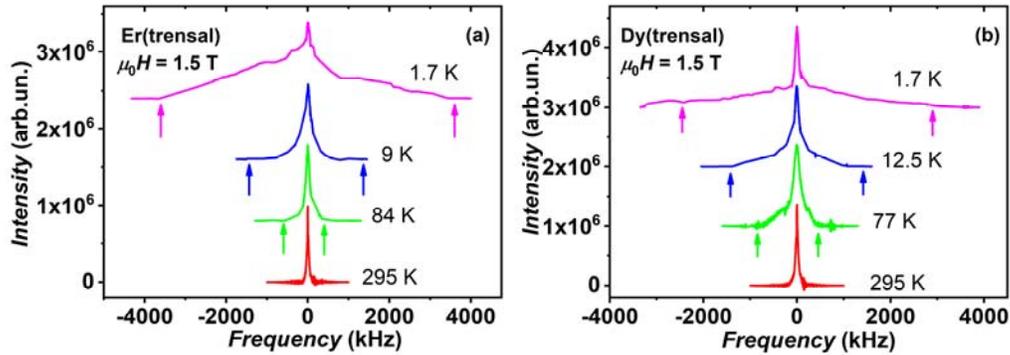

**Fig. 3** (color online) A collection of $^1$H NMR absorption spectra at different temperatures for Er(trensal) (a) and Dy(trensal) (b) samples in a magnetic field $\mu_0 H$ = 1.5 T. The arrows evidence the increase of the node-to-node width on decreasing temperature.

It is worth noting that for T>50K the NMR linewidth is proportional to the magnetic susceptibility. As the temperature is lowered below ~50K and the magnetic ground state becomes the only one populated, $\chi_{mol}T$ decreases (see Fig.2) and the NMR Full Width at Half Maximum (FWHM) of the central peak of the spectrum is no longer linear in the magnetic susceptibility. This behavior is highlighted in Fig. 4 for both derivatives (for applied field $\mu_0 H$ = 0.5 T). The temperature at which the deviation of FWHM from linearity in $\chi$ occurs corresponds to the temperature at which a change of slope in $\chi_{mol}T$ vs $T$ is observed in Fig. 2 (i.e. at about 50 K). It should be stressed that also the "node to node" width of the spectra plotted as a function of $\chi$ (data not shown) displays a departure from linear behavior at $T$~50K.

Finally, it can be noted that the "node to node" spectral width (reflecting the broad base behavior) increases further for T<15K, until at the lowest temperatures it reaches values as high as a few MHz.

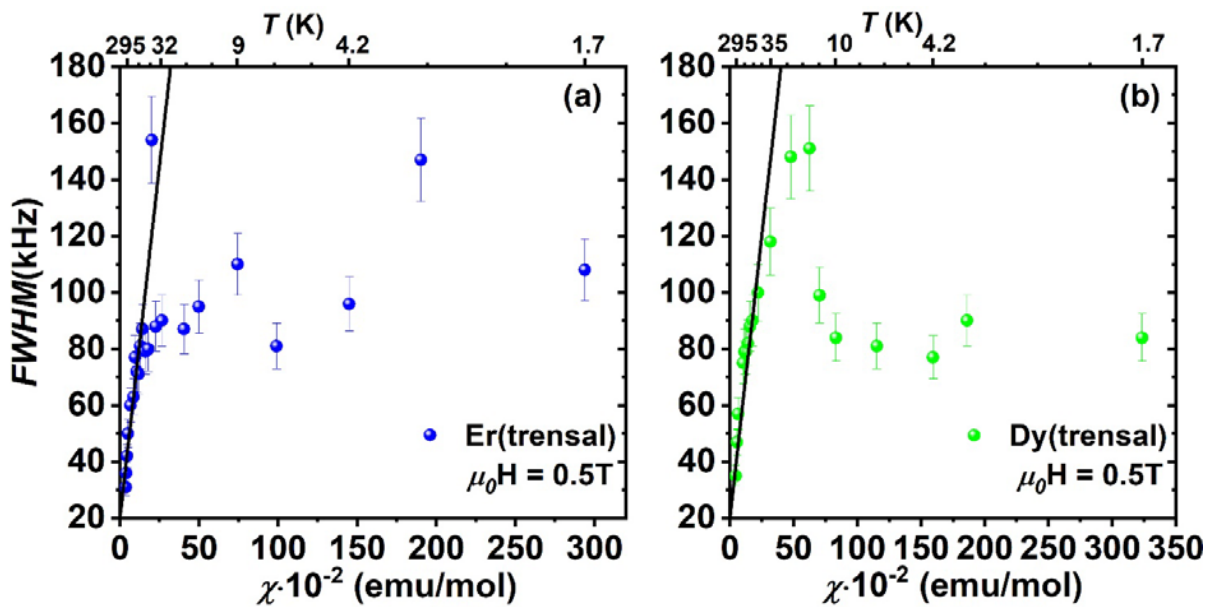

**Fig. 4** (color online) Full width at half maximum (FWHM) of the $^1$H NMR spectrum plotted as a function of the magnetic susceptibility for Er(trensal) (a) and for Dy(trensal) (b) at $\mu_0H = 0.5$ T. The black continuous lines evidence the regions where the NMR linewidths follow the linear behaviour expected for paramagnetic systems.

**V. Proton spin-lattice relaxation rate**

In order to get a more direct insight into the temperature dependence of the spin dynamics we performed proton spin-lattice relaxation rate measurements. All the recovery curves, plotted as $[1 - M_z(t) / M_z(\infty)]$ vs delay time (with $M_z(\infty)$ the longitudinal nuclear magnetization equilibrium value), resulted to have a bi-exponential behaviour in the entire temperature range investigated and for both applied magnetic fields (see Fig. S1 in the Supplemental Material [42]). The $1/T_1$ results shown here pertain to the fast component of the decay, dominant at high temperature and related to the protons closest, and thus more strongly coupled, to the magnetic moments of the lanthanide ion and strictly correlated to the spin dynamics of the electronic spin system. On decreasing temperature, the proton signal undergoes the so-called wipe-out effect [34b]: an increasing part of the nuclei does not contribute anymore to the NMR signal mainly because of the $T_2$ relaxation time shorteningThis effect is especially pronounced in the temperature range above the spin-lattice relaxation rate peak (see later), with a decrease of about 30 – 50% of the signal, depending on the field applied. It is however almost constant in the temperature range of the peak (see later). The wipe-out [34b] also leads to a change in the relative weights of the two components, with a decrease of the fast component. This however maintains a weight of about 30 – 50 % (depending on the sample and on the applied magnetic field). The fast decaying signal is then still reliable for our data analysis in the temperature range of the peak for both the samples investigated (see ref. [42], supplemental material)

The results of proton spin-lattice relaxation rate versus temperature at different fields are shown in Fig. 5 for the two samples investigated. In both systems the relaxation rate has a broad peak in the range 10 K < T < 50 K. Due to the relatively "high" weight of the fast component in the recovery curve, in principle the wipeout effect [34b] shown in Fig. 6 should not alter the analysis of the experimental results in a significant way since the variation of $M_{xy}(0)*T$ in the temperature region around the peak is small. However an effect of the wipeout on decreasing the absolute value of $1/T_1$ at all fields for T<100K cannot be completely excluded; this could alter the shape of the experimental curves.

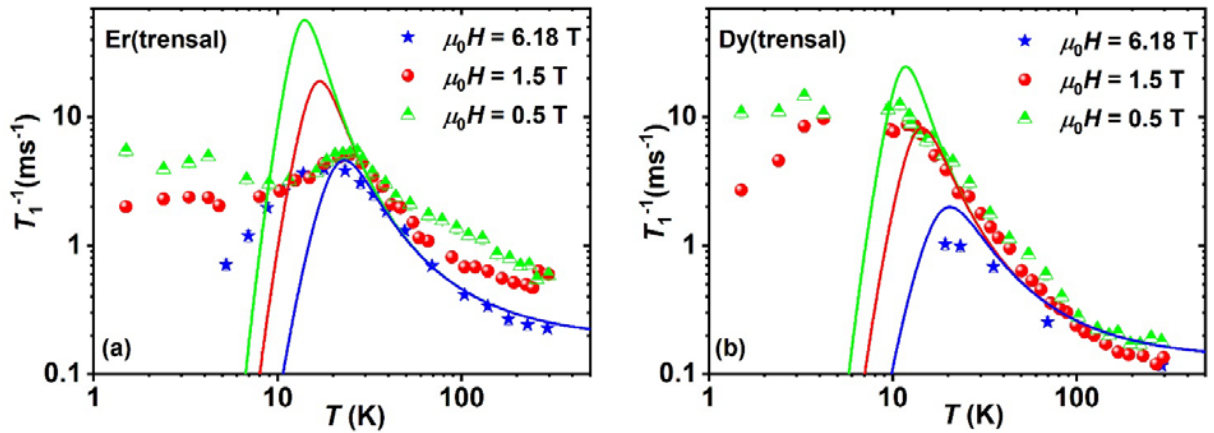

**Fig. 5** (color online) Proton spin/lattice relaxation rate vs. temperature at three different external magnetic fields for Er(trensal) (a) and Dy(trensal) (b). The full lines represent the behavior expected according to Eq. 2 as explained in the text.

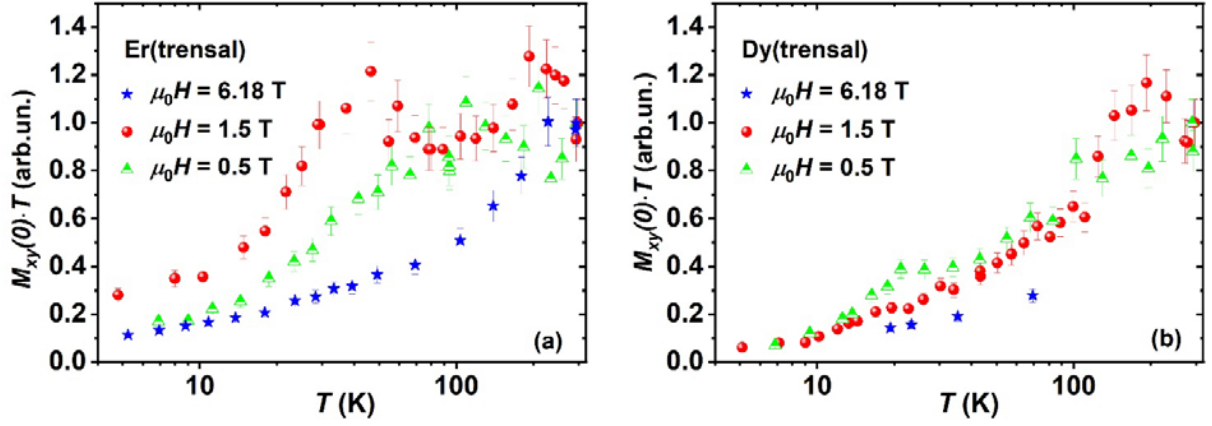

**Fig. 6** (color online) The quantity $M_{xy}(0) \cdot T$ reported as a function of temperature. $M_{xy}(0) \cdot T$ is proportional to the number of resonating nuclei and its decrease on temperature reflects the presence of the so-called wipeout effect [34b]. See text for details.

### VI. Muon spin lattice relaxation rate results

The muon asymmetry curves $A(t)$, see Figs. S2-S4 in Supplemental Material [42], were fitted with Mulab toolbox [43] and the muon relaxation rates λ were extracted and plotted as a function of temperature at different applied fields. For the fitting of the asymmetry relaxation curves, we used two different models: (i) a three-components fit, where we used the sum of three exponentials; (ii) a two-components fit, with an exponential plus a stretched exponential function. The behaviour of the longitudinal muon relaxation rate λ vs. T at different fields, is very similar for both models (see the comparison among Fig. 7 and Figs. S5-S6 in Supplemental Material [42]). Thus, it can be concluded that the information on physical properties extracted from μSR data are independent from the muon asymmetry fitting model. As the presence of three-components allows to obtain a better fitting (smaller errors, i.e. smaller $\chi^2$) and more detailed information on the muon polarization dynamics, in the following we will discuss the data on the basis of this model (see ref. [44] for a similar model).

The fitting function adopted in the entire temperature range was:

$$A(t) = a_1 \exp(-\lambda_1 t) + a_2 \exp(-\lambda_2 t) + a_3 \exp(-\lambda_3 t) + C_{bk} \qquad (1)$$

where $a_i$ represent the weights of the different exponentials and $\lambda_i$ the muon longitudinal relaxation rates. The $a_i$ values were obtained from an accurate comparison among the fitting results for high and low temperature data in different fields. This allowed to estimate (despite the 6-free parameters fitting by using Eq. 1) with reasonable precision their values, assuming that for each different compound $a_1$, $a_2$, $a_3$ are constant for any temperature and field. As usual, the values of $a_1$, $a_2$, $a_3$ reflect the relative percentage of muons implanting in at least three (the number of components) inequivalent sites. Moreover, since the two complexes are isostructural they should have the same weights for the three components (since the muon should implant at the same sites in the two samples). This means also that, once the value of the initial asymmetry is identified, each component should represent the same percentage of the total asymmetry in both compounds. Best fit results were then obtained with the following parameters: $a_1 = 0.1$, $a_2 = 0.08$, $a_3 = 0.05$, and the background contribution coming from sample holder was estimated to be $C_{bk} = 0.02$ for Dy(trensal) and $C_{bk} = 0$ for Er(trensal).

It is worth stressing that:

(i) the fastest relaxing component $a_1$, pertaining to muons implanting closest to the magnetic centres (and thus with the strongest magnetic interaction), is characterized by a very fast relaxation rate $\lambda_1$, especially for temperatures T < 30 K, which is responsible for the drop of the total muon asymmetry (see Figures S3 and S4) at very short times and low temperature. However, due to frequency window limitation imposed by the μ$^+$SR technique, $\lambda_1$ values are often too high ($\lambda_1 > 20$ μs$^{-1}$, see Fig. S7 in Supplemental Material [42]) and, as a consequence, the related curves mostly unreliable;

(ii) the slowest relaxation rate $\lambda_3$ has a behaviour vs T and H similar to the ones of the "intermediate" rate $\lambda_2$ (see Fig. S8 in Supplemental Material [42]), but pertains to muons less coupled to the magnetic ions. It is thus less informative for what concerns the molecular spin dynamics. Additionally $a_3<a_2$, thus giving bigger fitting errors.

We then decided to focus on the data obtained for the component presenting intermediate values of relaxation rate, $\lambda_2$. The temperature dependences for both derivatives at different fields are shown in Fig. 7. It is evident that the relaxation rates present a peak at around 10-20 K which shifts at higher temperatures and increases in amplitude with increasing the external applied magnetic field. We stress again that, qualitatively, this result is independent of the model used to fit the asymmetry curves and can be deduced with a direct by-eye analysis of the raw data (Fig. S3 in Supplemental Material [42]). Indeed, asymmetry curves measured at the same temperature (close to 10 K) and different fields clearly points to a slower relaxation at lower magnetic field. At the same time, for temperatures lower than 5 K, the relaxation is clearly temperature independent (Fig. S4 in Supplemental Material [42]).

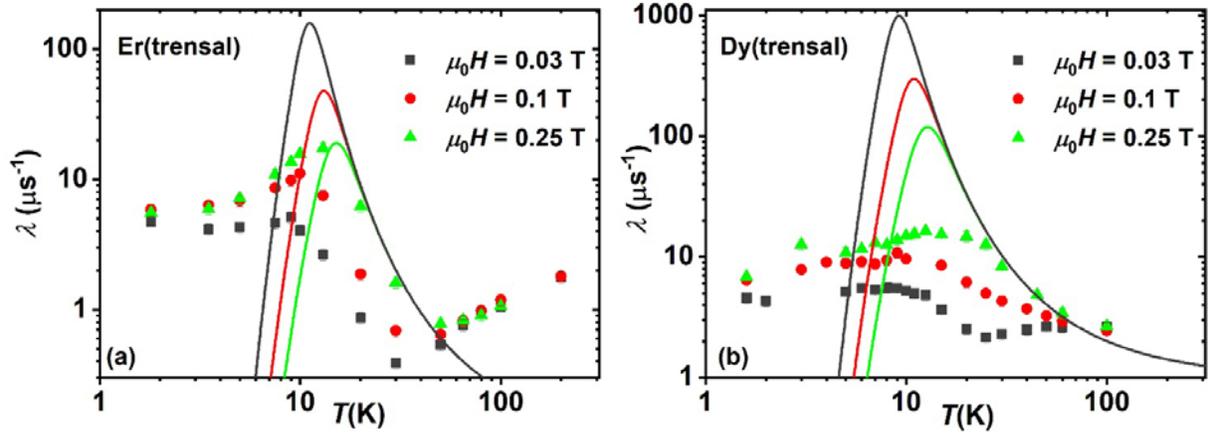

**Fig. 7** (color online) Muon spin-lattice relaxation rate vs. temperature at three different external longitudinal magnetic field for Er(trensal) (a) and Dy(trensal) (b).

### VII. Discussion

The results reported in the previous paragraphs indicate that a broadening of the NMR line $\Delta\nu$ (up to 0.5 MHz and more) at temperatures as high as 50-60 K was detected. Furthermore, a clear peak was observed at all applied magnetic fields in the longitudinal relaxation rates $\lambda$ ($\mu^+$SR, peaks in the region 10<T<20K, depending on the field and the Ln ion) and $1/T_1$ (NMR, peaks in the region 10<T<30K). In general terms, the NMR line broadens when the characteristic correlation rates are of the order of the linewidth $2\pi\Delta\nu = \Delta\omega$ or smaller, while the relaxation rates may show a peak when the correlation rates are of the order of the Larmor frequency).

The observation of a peak in the relaxation rate is a common occurrence for all the molecular magnets investigated previously by $^1$H NMR or $\mu^+$SR [23, 26-29,45-49]. For those systems, the $1/T_1$ (or $\lambda$) vs T plot could be fitted well by an expression derived from the general formula of Moriya for nuclear relaxation in paramagnets [30,31] (BPP function) based on the presence of a single correlation frequency:

$$1/T_1 \text{ (or } \lambda\text{)} = A\,\chi T \omega_c / (\omega_c^2 + \omega_L^2) \quad (2)$$

where $\chi T$ is dimensionless, $\omega_L$ is the Larmor frequency of the nucleus (muon), A is the strength of the geometric part of the hyperfine interaction and $\omega_c$ the characteristic correlation frequency of the magnetic fluctuations. If both A and $\omega_c$ are magnetic field independent, Eq. 1 predicts that the amplitude of the peak of the relaxation rate should scale as $1/\mu_0 H$ and that the peak should move to lower temperature as the field decreases (for the usual case of a slowing down of the spin fluctuations on lowering the temperature). Further, Eq. 1 predicts that in the fast motion regime (i.e.

$\omega_c \gg \omega_L$) relaxation rates should be field independent. Even for cases where approaches alternative to BPP have been used to interpret the nuclear (muon) spin-lattice relaxation in molecular magnets, the expression for the relaxation rate is similar to Eq. 1 and the same scaling behavior should be observed in case of field independent parameters.

It is quite evident from both Fig. 5 and Fig. 7 that the present results cannot be interpreted in terms of Eq. 1, differently from what found in all other molecular nanomagnets previously investigated by these techniques. [23, 26-29,45-49] Indeed, the dependence of the height of the peak upon external magnetic field is opposite to the one predicted by Eq. 1. This qualitative analysis is confirmed by plotting the behavior expected for Eq. 1 and assuming an Arrhenius law for the temperature dependence of the correlation frequency, i.e. $\omega_c = \omega_0 \exp(-\Delta/T)$. Here $\Delta$ is the height of the thermal activation barrier, which in the case of SMMs is related to the magnetic anisotropy, and $\omega_0$ is the correlation frequency at infinite temperature. The theoretical curves in Fig. 5 and 7 were obtained by using in Eq. 1 the experimental $\chi T$ values, $\omega_0 = (6.5 \pm 3) \cdot 10^{10}$ s$^{-1}$ $\Delta = (77 \pm 2)$ K for Er(trensal), and $\omega_0 = (5 \pm 2) \cdot 10^{10}$ s$^{-1}$ $\Delta = 62 \pm 10$ K for Dy(trensal). The $\Delta$ value was chosen to be the energy of the first excited state for both systems, as obtained by other theoretical and experimental techniques [35, 37, 39]. We note here that while ac susceptibility data could be modeled by including Raman and direct relaxation processes, these were found to be relevant in a much lower temperature region with respect to the one of the peak and were thus not included here. Anyhow, also the inclusion of all the relaxation processes in $\omega_c$ does not allow to match the experimentally observed field and temperature dependence of $1/T_1$ (see Fig. S9 in Supplemental Material [42]).

Furthermore, it is worth mentioning that even assuming a distribution of correlation times in place of a single one, the data fitting cannot be improved significantly. Indeed, the presence of a distribution affects only the BPP function on the left side of the peak and just slightly changes the ratio among $1/T_1$ peaks at different fields.

In the absence of an appropriate quantitative theoretical model to rationalize the observed results, which is beyond the scope of this article, we propose here a simple qualitative model which may explain the results of both NMR and $\mu^+$SR relaxation experiments. As discussed above, the FWHM of the central peak of the NMR spectrum is no longer linear in the magnetic susceptibility below about 50 K, indicating a slow dynamics with a characteristic correlation frequency of few MHz or less already at 50 K. Conversely, the observed peaks in $1/T_1$ and $\lambda$ point to dynamics on the scale of tens/hundreds of MHz at lower temperatures (10-20 K). Thus, we suggest that in these systems, two different independent relaxation dynamics are active: the first dominating for T>50K and characterized by a correlation time $\tau_{HT}$, and the second dominating for T<50 K, characterized by a correlation time $\tau_{LT}$. This hypothesis is suggested by combining the general temperature dependence of the longitudinal muon and proton relaxation rates of the systems with the behavior of the NMR line width as a function of temperature. In particular, the presence of a slow component ($1/\tau_{HT}$ ~ few MHz) of the spin dynamics already at high temperature is consistent with the field dependence of the nuclear relaxation rate $1/T_1$ (Fig. 5) for T > 50-60 K. This behavior could be generated by the slow motion of the spins when the dominating rate is $1/\tau_{HT}$, i.e. for $1/\tau_{HT} << \omega_L$. This slow motion can be responsible of the typical field and temperature dependence of a $1/T_1$-BPP function (Eq. 1, where the correlation frequency is $\omega_{c\text{-high}} = 1/\tau_{HT}$), whose peak occurs at T > 300 K. Thus, while the peak itself is outside the investigated temperature range, the peculiar behaviour we observe has to be traced back to the low temperature tail of the (T>300K) peak. As discussed above, in the contrasting hypothesis of fast dynamics, i.e. a correlation frequency much larger than the Larmor frequency ($1/\tau_c >> \omega_L$), $1/T_1$ would be field independent. This possibility has then to be discarded.

At lower temperatures (about 10-20 K), where the susceptibility becomes field dependent (see Fig. 2), a faster spin dynamics (determined by $1/\tau_{LT}$, of the order of 10-100 MHz) sets in and the system condenses in the doublet ground state that behaves as a "frozen" spin state. For the onset of a frozen spin state there are two possible scenarios:

(i) a long-range 3D magnetic ordering among the single ion molecules occurs, and generates a peak in the relaxation rate at the transition temperature [50]. Indeed, the relatively large Ln-Ln intermolecular closest distance (7.69 Å) and the absence of intermolecular superexchange paths excludes the possibility of long range order in the investigated temperature range due to the (exchange and) super-exchange interaction, and points to a purely molecular origin of the observed slow relaxation. The latter was indeed observed by ac susceptometry even in samples diluted in an isostructural diamagnetic matrix.[37] Accordingly, sample calculations provide an estimated value of some tenth of gauss for the dipolar interaction acting among the RE ions of the different molecules (see supplemental material [42] for details), thus suggesting an ordering temperature below 1 K i.e. outside our experimental data range;

(ii) a short range continuous freezing of the moments of the RE ion related to the gradual occupation of the ground state doublet is in order. The magnetic moments of the complexes should be considered frozen when their fluctuation frequency becomes smaller than the characteristic frequency of the hyperfine interactions, which are of the order of a few hundred of KHz/ few MHz. In this scenario, the fluctuations of the magnetization involve a large variation of the local field at the nuclear (muon) site and the corresponding relaxation rate originates from the direct exchange of energy among the $^1$H nuclear (muon) levels and the electronic molecular levels broadened by the hyperfine and/or the intermolecular dipolar interactions.[51] In order to analyze quantitatively the relaxation data in this low temperature range a detailed theoretical model, outside the scope of the present paper, is required.[1]

Finally, in the lowest investigated temperature region (1-3 K) both the NMR and the $\mu^+$SR results, shown in Fig. 5 and 7, respectively indicate that the relaxation rates tend to become temperature independent, thus suggesting the presence of a quantum phenomenon.

As final further tentative of quantitative data rationalization, taking into account the experimental data and the above discussion, we setup a phenomenological model by assuming two additional hypotheses: (i) in Eq.(2), $1/T_1$ (or $\lambda$) is not assumed to be proportional to $\chi T$. As a consequence, the hyperfine coupling A in eq.(2) could be field-dependent; (ii) at T<4-5K, the dominating correlation time becomes temperature-independent, as suggested by the behavior of the NMR and $\mu^+$SR relaxation rates, while maintaining the field dependence. We stress that the NMR data can be affected by the wipeout effect that could alter significantly the $1/T_1$ amplitude at all fields, and so in the following discussion they will not be taken into account.

With the above hypotheses, the expression of the muon longitudinal relaxation rate becomes:

$$\lambda = A(H) \omega_{c1} / (\omega_{c1}^2 + \omega_L^2) \quad (3)$$

where A(H) is an effective field-dependent hyperfine coupling, and the correlation frequency $\omega_{c1}$ is written as : $\omega_{c1}(H) = \omega_{01} \exp(-\Delta/T) + \omega_T(H)$. In this case $\Delta$ is assumed to have the same values reported above for the two systems, $\omega_{01}$ is the usual tentative frequency and $\omega_T(H)$ is the field dependent term of quantum origin, whose value does not change with temperature. As can be seen from Fig. S10 in Supplemental Material [42], the agreement with the $\mu^+$SR experimental data

---

[1] In this situation, the weak collision approach is no longer valid; in absence of an external magnetic field or if the applied field is negligible compared to the internal local static fields a strong collision process might be considered. However, this regime can be excluded here since the measurements were performed in an external magnetic field which possibly is comparable to the internal local field

improves for T < $T_{peak}$, while for T > $T_{peak}$ in most cases the agreement is poor. On the other hand, the data fit for T > $T_{peak}$ could be significantly improved by taking into account the contribution of the processes characterized by the correlation time $\tau_{HT}$, whose quantitative parameters are however unknown.

It should be finally remarked that the field and temperature dependence of the relaxation time of the magnetization were previously reported [37] in the same systems, as obtained from susceptibility measurements. However, in comparing the NMR ($\mu^+$SR) spin dynamics results with the susceptibility data, one should be aware that the macroscopic relaxation time of the magnetization measures the spin fluctuations of the q = 0 mode, while the microscopic [50]. It is, however, significant that both the microscopic and the macroscopic techniques show that the relaxation time of the magnetization becomes temperature independent at low temperature, indicating that the dominant relaxation mechanism is of quantum nature.

## IV. Summary and conclusions

We presented proton NMR and $\mu^+$SR measurements over a wide temperature and magnetic field range in two Lanthanide based molecular magnets, namely Er(trensal) and Dy(trensal).

Our experimental results highlight an unconventional spin dynamics in the two complexes investigated. For both molecular systems, at temperatures of the order of 50K the $^1$H NMR spectrum starts to broaden well outside the typical paramagnetic effect. This indicates that, in the high temperature region, the dynamical magnetic fluctuations are dominated by a correlation frequency $1/\tau_{HT}$ that becomes of the order of the spectral width, i.e. some hundreds of kHz/ few MHz, for T~50K. This dynamics could be related to the spin relaxation of magnetic excited states and could became unimportant at low temperatures, when only the electronic ground doublet is populated. On the other hand, at lower temperatures (about 10-25K, depending on the compound), the nuclear $1/T_1$ and muon $\lambda$ spin-lattice relaxation rates exhibit a peak which cannot be associated to the slowing down of the dynamics dominating at high T, because the related Larmor frequency $\omega_L$ is of the order of tens/hundreds of MHz, i.e. much higher than $1/\tau_{HT}$ and so outside the resonance condition $\omega_L \cdot \tau_{HT} \approx 1$. Thus, as the full experimental results cannot be explained in terms of a dominating single correlation frequency which decreases as the temperature is lowered, we propose the insurgence of two independent spin dynamics in different temperature ranges, T>50K and T<50K. The longitudinal relaxation rate peak has thus been attributed to the insurgence of a ground state spin dynamics whose correlation frequency was called $1/\tau_{LT}$. Finally, for T<4-5K we observed a flattening of the spin-lattice muon and nuclear relaxation rates, particularly evident at the lowest fields, possibly of quantum nature and related to the magnetization tunneling.

By concluding, the present results are of relevance since they contrast with those hitherto reported for lanthanide based molecular complexes, which were amenable to the simple model of the slowing down of a single correlation time (the well-known Bloembergen-Purcell-Pound, BPP, model), and can provide further information on the microscopic details of the relaxation processes in some of these systems. Further examples of a behaviour similar to the one reported here are expected to arise in the next future, given the huge interest in the spin dynamics of these molecules. In perspective, this will require the development of an appropriate theoretical modeling going beyond the simple qualitative understanding. This is of particular importance for NMR and $\mu^+$SR data, since their sensitivity makes them techniques of choice to study the spin dynamics at the nanoscale [52].

### Acknowledgments

We acknowledge the financial support of MIUR through the project Futuro in Ricerca 2012 (RBFR12RPD1). S. C. and L. S. also acknowledge support from the European Project SUMO of the call QuantERA 2017.